%% file: EuroGNC2024.tex
\documentclass{CEAS_EuroGNC_2024}

\usepackage[utf8]{inputenc}

\usepackage{graphicx}
\usepackage{amsmath}
\usepackage[version=4]{mhchem}
\usepackage{siunitx}
\usepackage{array}
\usepackage{longtable}
\usepackage{wrapfig}

\usepackage{url,hyperref,lineno,microtype,subcaption}
\usepackage{algorithm} 
\usepackage{algpseudocode}
\usepackage{bm}
\usepackage{multirow}
\usepackage[english]{babel}
\addto\extrasenglish{%

}
\usepackage[todonotes={textsize=tiny, disable},commentmarkup=todo, final]{changes}

\title{Region of Attraction Estimation for Free-Floating Systems under Time-Varying LQR Control}

\ifpdf
\hypersetup{pdfinfo={
Title={Region of Attraction Estimation for Free-Floating Systems under Time-Varying LQR Control},
Author={Shubham Vyas},
Subject={Region of Attraction Estimation for Free-Floating Systems under Time-Varying LQR Control},
Keywords={Active Debris Removal; ADR; Time-Varying Linear Quadratic Regulator; TVLQR; regions of attraction; ROA; free-floating robot; space robot; space robot control; space robot experimental validation}
}}
\fi

\begin{document}
\maketitle

\begin{authorList}{4cm}
\addAuthor{Lasse Shala}{Researcher, Robotics Innovation Center, German Research Center for Artificial Intelligence (DFKI GmbH), 28359 Bremen, Germany.}
\addAuthor{Shubham Vyas}{Researcher, Robotics Innovation Center, German Research Center for Artificial Intelligence (DFKI GmbH), 28359 Bremen, Germany. \emailAddress{Shubham.Vyas@dfki.de}}
\addAuthor{Mohamed Khalil Ben-Larbi}{Postdoctoral Fellow and Group Lead, Chair of Space Technology, TU Berlin, 10587 Berlin, Germany.}
\addAuthor{Shivesh Kumar}{Researcher, Robotics Innovation Center, German Research Center for Artificial Intelligence (DFKI GmbH), 28359 Bremen, Germany. }
\addAuthor{Enrico Stoll}{Full Professor, Chair of Space Technology, TU Berlin, 10587 Berlin, Germany.}
\end{authorList}

\begin{abstract}Future Active Debris Removal (ADR) and On Orbit Servicing (OOS) missions demand for elaborate closed loop controllers. Feasible control architectures should take into consideration the inherent coupling of the free floating dynamics and the kinematics of the system. Recently, Time-Varying Linear Quadratic Regulators (TVLQR) have been used to stabilize underactuated systems that exhibit a similar kinodynamic coupling. Furthermore, this control approach integrates synergistically with Lyapunov based region of attraction (ROA) estimation, which, in the context of ADR and OOS, allows for reasoning about composability of different sub-maneuvers. In this paper, TVLQR was used to stabilize an ADR detumbling maneuver in simulation. Moreover, the ROA of the closed loop dynamics was estimated using a probabilistic method. In order to demonstrate the real-world applicability for free floating robots, further experiments were conducted onboard a free floating testbed.
\let\thefootnote\relax\footnotetext{Corresponding Author: Shubham Vyas \emailAddress{Shubham.Vyas@dfki.de}}
\end{abstract}




\section{Introduction}
\input{sections/introduction}

\section{Background}
\input{sections/background}

\section{Region of Attraction Estimation}
\input{sections/roaest}

\section{Results}
\input{sections/results}

\section{Discussion}
\input{sections/discussion}

\section{Conclusion}
\input{sections/conclusion}

\section*{Acknowledgments}
This work has been performed in the M-RoCK project funded by the German Aerospace Center (DLR) with federal funds (Grant Number: FKZ 01IW21002) from the Federal Ministry of Education and Research (BMBF) and is additionally supported with project funds from the federal state of Bremen for setting up the Underactuated Robotics Lab (Grant Number: 201-001-10-3/2021-3-2).
The second author acknowledges support from the Stardust Reloaded project which has received funding from the European Union’s Horizon 2020 research and innovation programme under the Marie Skłodowska-Curie grant agreement No 813644.

The authors would also like to acknowledge the ELISSA Lab at the Institute of Space Systems, TU Braunschweig, for providing the facilities and support that contributed significantly to the results presented in this paper. The research was conducted during the third author's time at TU Braunschweig, who is currently affiliated with the SmallSat Rendezvous \& Robotics Group at the Chair of Space Technology, TU Berlin.

\bibliography{references}

\end{document}

%% file: sections/introduction.tex
\label{text:intro}
Future spaceflight endeavors, undertaken by government agencies and private companies, envision a sustainable use of Low and Geostationary Earth Orbit (LEO/GEO) and further venturing out into the solar system. May it be the life extension vehicles by Northrop Grumman~\citep{Redd.2020}, manned missions to Mars by SpaceX~\citep{Heldmann.2022}, or the international Lunar Gateway project~\citep{Burns.2019} involving NASA, ESA, JAXA and CSA, a common challenge for many upcoming projects is the in-orbit operations process. This will require capabilities to manipulate and maneuver accurately with and around cooperative and non-cooperative objects in orbit. 

Active debris removal (ADR) and on-orbit servicing (OOS) are required to reduce or at least stabilize the current space debris environment~\cite {BenLarbi.2022, BenLarbi.2017}. In such a mission an active spacecraft (Chaser) would approach a debris object (Target), dock to it (e.g. using a robotic manipulator), optionally detumble, and finally remove it from the region of concern (ADR) or extend its operational life (OOS) through servicing measures, i.e. refueling or repair. This will help to prevent collisions that would otherwise render parts of the orbit inoperable for decades to come~\citep{BenLarbi.2022}. 
Furthermore, projects involving large structures in space require powerful and dedicated launchers, which constitute a technical and financial challenge. However, this limitation can be overcome by innovative techniques, such as additive manufacturing, which makes it possible to build components and assemble them directly in space, putting virtually no constraints on the structure's geometry. 

\begin{wrapfigure}{r}{0.3\linewidth}
    \centering
    \includegraphics[width=\linewidth]{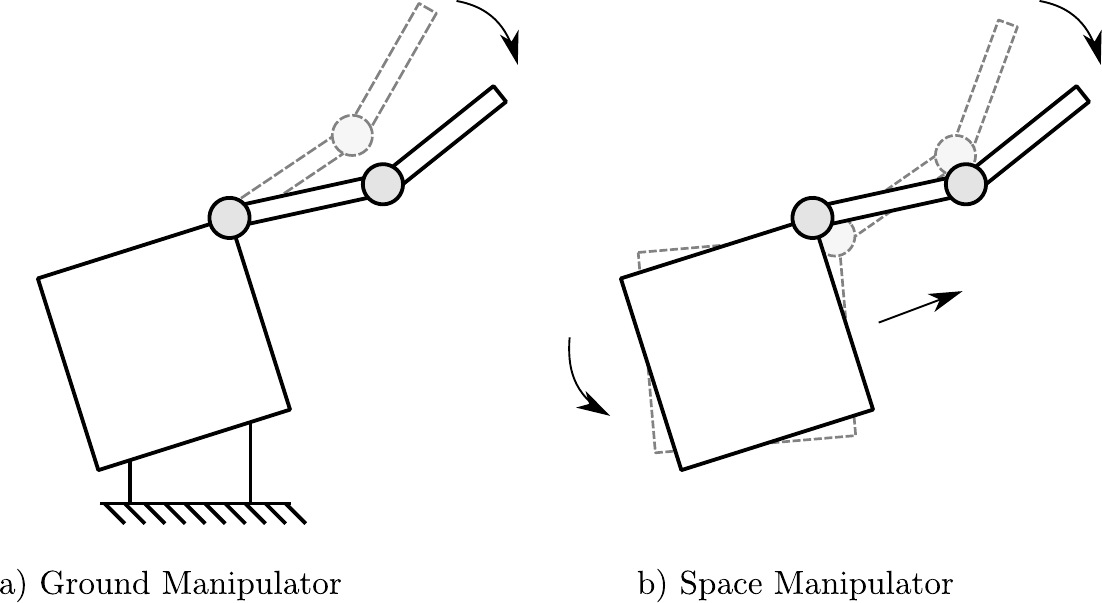}
    \caption[Coupling of kinematics and dynamics.]{Coupling of kinematics and dynamics, adopted from \cite{xu1993}. For manipulators fixed to the ground, there is no significant motion of the base link, when moving individual joints. The contrary holds for manipulators mounted on a spacecraft.}
    \label{fig:couplingKinDyn}
\end{wrapfigure}

For all the aforementioned applications, sophisticated control systems are required to take care of the simultaneous movement of a mobile base, i.e. the spacecraft, and a robotic arm (potentially grasping a target object) to allow for efficient manipulation, flawless assembly, or continuous printing. 
For a free-floating manipulator, any end effector position change in an inertial frame is a function of the change in the joint positions as well as the configuration-dependent inertia properties~\citep{FloresAbad.2014}. Therefore, in general, the inverse and forward kinematics for space manipulators are also considered dynamics problems~\citep{FloresAbad.2014}. This kino-dynamic coupling is illustrated in \autoref{fig:couplingKinDyn}. Moreover, the time-varying motion constraints imposed by a moving, potentially tumbling, target further complicates the grasping task and requires agile planning capabilities within an operationally useful time frame.

\begin{wrapfigure}{r}{0.6\linewidth}
\centering
\includegraphics[width=\linewidth]{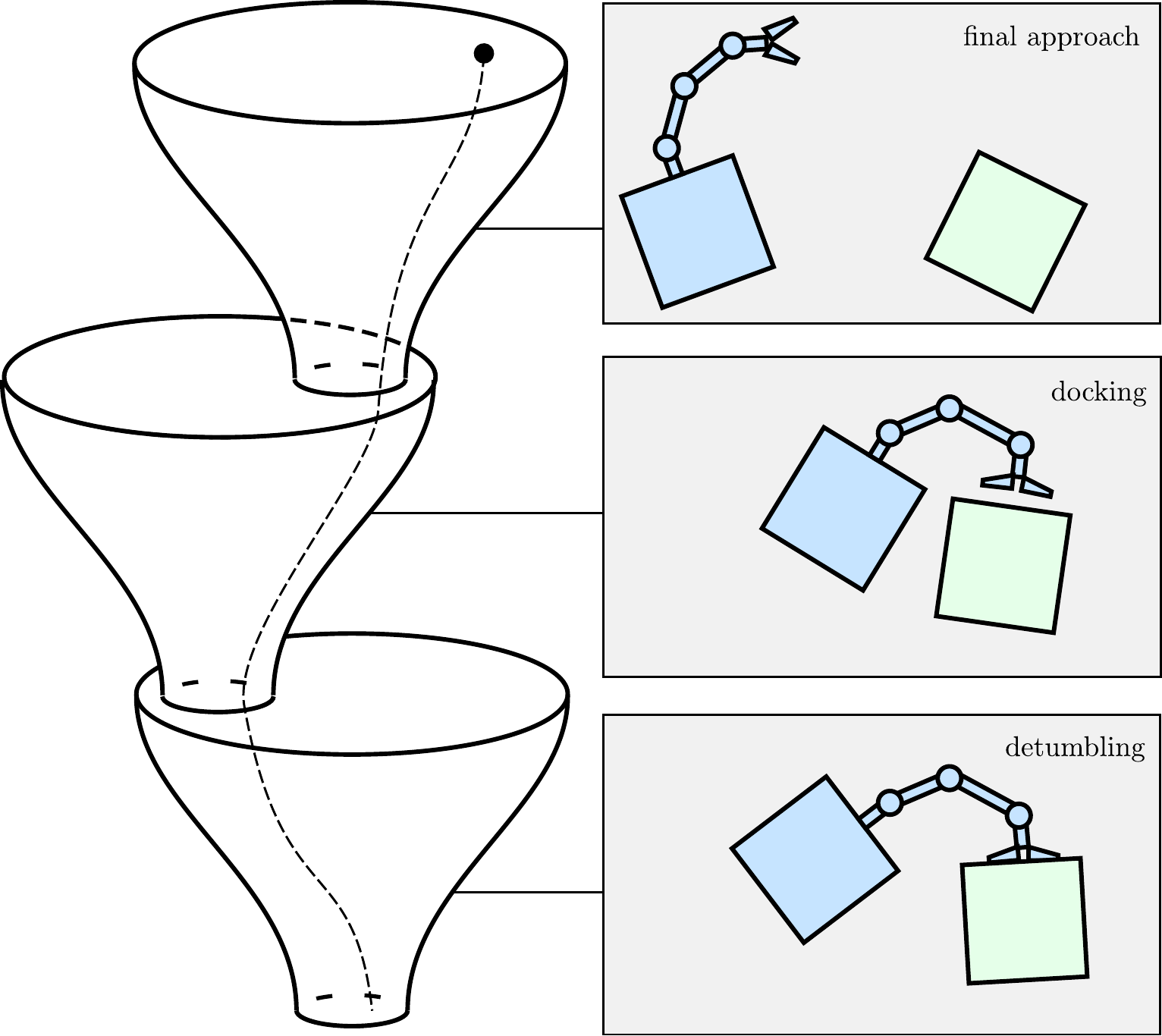}
\caption{Composition of a multi-phase ADR scenario using funnels in the spirit of \citep{burridge1999}. Each funnel represents the stabilizing closed-loop dynamics during a given phase of ADR. Stability over the entire maneuver can be achieved by ensuring that states at the outlet of a preceding funnel are a subset of the states of the subsequent one.}
\label{fig:funnel_motivation}
\end{wrapfigure}

While real-time dynamic re-planning is an increasingly plausible alternative \citep{virgili-llop2019, rybus2017, albee2019}, the majority of the proposed techniques still face significant performance challenges because of the high computational requirements for on-board processors and the inability to convexify certain constraints~\citep{virgili-llop2019}. 
To address these issues, the control problem can be split into two parts. In this case, the trajectory optimization problem is solved offline, or at a non-real-time rate, and a linear controller is used for online stabilization. 
Within the field of optimal control of underactuated systems, the usage of Time-Varying Linear Quadratic Regulators (TVLQR) as locally optimal feedback control policy to stabilize planned trajectories has gained popularity~\citep{underactuated,reist2010,reist2016,moore2014,majumdar2017}. TVLQR requires little computational capacity, can be run at high frequencies, and integrates synergistically with Lyapunov-based methods for Region of Attraction (RoA) estimation that can be used to verify closed-loop stability. So-called "funnels" describe time-varying regions of certifiable stability around trajectories~\cite {tedrake2010,majumdar2017,tobenkin2011}. In the context of ADR and OOS, such analyses are especially interesting when considering space qualification of control algorithms.  
Furthermore, RoA analysis could be an interesting tool enabling controller composition in the spirit of \cite{burridge1999}, where it could be used to allow for safe transitioning from one phase of a maneuver to the next. 
For example, considering a robotic ADR scenario consisting of a final approach, capture, and a post-capture/detumble phase~\citep{flores-abad2014}, determining the set of stabilizable states of the closed-loop dynamics during each phase is beneficial.
Doing so allows the connection of each individual phase to the subsequent one while ensuring the stability of the entire ADR maneuver (depicted in \autoref{fig:funnel_motivation}).
Within this work, a probabilistic RoA estimation pipeline for free-floating spacecraft under TVLQR control was conceived and applied to two distinct scenarios, namely: \textit{scenario I}, a simulated multibody scenario, and \textit{scenario II}, a simple trajectory tracking scenario that was realized onboard the ELISSA free floating testbed~\citep{Trentlage.2018c, Yang.2021}. Within \textit{scenario I}, the goal was to investigate the usefulness of probabilistic RoA estimation in an ADR context, whereas for \textit{scenario II}, the focus was put on investigating the real-world applicability of RoA estimates. 

This paper is organized as follows: \autoref{text:background} provides background information on the dynamical models and the control approach used throughout this work and shortly introduces the concept of RoA. 
In particular, the control approach used within this work consists of stabilizing a precomputed nominal open loop trajectory obtained using trajectory optimization (\autoref{subsec:trajopt}) with TVLQR (\autoref{text:stabilization}).
This method has proven to be capable of stabilizing a range of systems with similar kinodynamic couplings, i.e. severely underactuated systems \citep{moore2014,cory2010}, fully-actuated \citep{Vyas2022}, and 3-DoF freeflyers \citep{bredenbeck2022}. 
\autoref{text:roa_est} introduces the probabilistic RoA estimation methodology that is then applied to the two aforementioned scenarios (\autoref{text:results}). The methodology and results are then discussed in \autoref{text:discussion}.

%% file: sections/background.tex
\label{text:background}
\subsection{Equations of Motion}
The equations of motion (EoM) for free-floating multibody systems have been studied in detail in the literature. Hence, we only provide the relevant equations here. For in-depth information on the derivation of the EoM, we direct the reader to the following works: \cite{DubowskySteven1993, Papadopoulos1990a, Wilde2018a, NguyenHuynh2013}. 
\replaced[id=LS, comment=2]
{With $\mathbf{q} = \begin{bmatrix} \bm q_0^T \ \ \  \bm q_m^T \end{bmatrix}^T$ 
denoting the pose of the freeflyer that is composed of the attitude of the base and the joint angles 
of the manipulator, respectively, the EoM can be written as}
{We can use the general multibody dynamics equations to write the EoM as}
(\cite{LynchPark2017}):

\begin{equation}
\label{eq:dynamicseqn}
\mathbf{M}(\mathbf{q}) \ddot{\mathbf{q}} +\mathbf{C}(\mathbf{q}, \dot{\mathbf{q}}) = \mathbf{u}
\end{equation}

Here, $\mathbf{q}$ and $\dot{\mathbf{q}}$ are the generalized positions and velocities of the system, respectively, $\mathbf{M}\in \mathbb{R}^{6+n \times 6+n}$ is the
\added[id=LS, comment=3]{invertible}
generalized mass-inertia matrix of the system, $\mathbf{C}(\mathbf{q}, \dot{\mathbf{q}})\in \mathbb{R}^{6+n}$ is the generalized Coriolis and Centrifugal effort, and $\mathbf{u}\in \mathbb{R}^{6+n}$ is the vector of generalized forces with $n$ being the number of degrees of freedom in the robot arm attached to the base. The equations of motion can also be represented in state-space form by defining $\mathbf{x} = [\mathbf{q}, \dot{\mathbf{q}}]^T$. Doing so, the derivatives of $\mathbf{x}$ can be written as:

\begin{equation}
\dot{\mathbf{x}} = \begin{bmatrix}
    \dot{\mathbf{q}} \\ \ddot{\mathbf{q}}
\end{bmatrix}    = 
\begin{bmatrix}
    \dot{\mathbf{q}} \\ \mathbf{M}(\mathbf{q})^{-1} [\mathbf{u} - \mathbf{C}(\mathbf{q}, \dot{\mathbf{q}}]
\end{bmatrix} 
\end{equation}

The mass inertia matrix of the system can be written as:

\begin{equation}
    \mathbf{M} = \begin{bmatrix}
    \mathbf{M_0} &  \mathbf{M_{bm}} \\
    \mathbf{M_{bm}} & \mathbf{M_{m}}
    \end{bmatrix}
\end{equation}

where $\mathbf{M_0} \in \mathbb{R}^{6 \times 6}$ is the base-spacecraft mass-inertia matrix component, $\mathbf{M_m} \in \mathbb{R}^{n \times n}$ is the manipulator mass-inertia matrix component, and $\mathbf{M_{bm}} \in \mathbb{R}^{6 \times n}$ is the dynamic-coupling inertia matrix component (\cite{Wilde2018a}). 
\added[id=LS,comment=4]{
Note, that in \autoref{subsec:detumbleresults}, the mass inertial properties
of the debris are also modeled in $\mathbf{M_m}$.}
Using the components of the mass-inertia matrix, the linear and angular momentum of the free-floating system can be written as:

\begin{equation}
\label{eq:momentumeqn}
\begin{bmatrix}
    \mathbf{L} \\ \mathbf{P} 
\end{bmatrix} 
= \mathbf{M_0}\dot{\mathbf{q}}_0 +  \mathbf{M_m}\dot{\mathbf{q}}_m
\end{equation}

Here, $\mathbf{L} \in \mathbb{R}^3$ and $\mathbf{P} \in \mathbb{R}^3$ represent the total angular and linear momentum of the system respectively. $\dot{\mathbf{q}_0} \in \mathbb{R}^6$ and $\dot{\mathbf{q}_m} \in \mathbb{R}^n$ are the generalized velocities of the base and manipulator respectively. The angular momentum equations form a non-holonomic constraint on the system which makes the control of the system complex.

\subsection{Trajectory Optimization}
\label{subsec:trajopt}
For every phase of ADR, a nominal state-space trajectory of the system has to be found which accomplishes the task for the given phase. In addition to being able to realize the phase's task, the trajectory also has to be dynamically feasible i.e. satisfy the dynamics constraints, minimize certain costs, and also allow for actuation and state constraints. Trajectory optimization based on Direct Transcription \citep{Betts2010, Kelly2017} was employed to find such a trajectory. The general formulation of the trajectory optimization problem can be written as:

\comment[id=LS]{Changed min to argmin in eq. 5a (comment 7)}
\begin{subequations}
\label{eq:trajopt}
\begin{equation}
\begin{aligned}
\label{eq:costfunctrajopt}
\mathbf{q}\left( t\right), \dot{\mathbf{q}}\left( t\right) ,\mathbf{u}\left( t\right) = \text{arg min}
& \int ^{t_{f}}_{t_0} \left( w_t \Delta t + \mathbf{u}^T \mathbf{W}_u \mathbf{u} \right) dt + \mathbf{x}_f^T \mathbf{W}_{\mathbf{x_f}} \mathbf{x}_f \\
\textrm{such that:} 
\end{aligned} 
\end{equation}
\begin{equation}
\label{eq:trajoptdyneqn}
\mathbf{M}(\mathbf{q}) \mathbf{\ddot q} + \mathbf{C}(\mathbf{q},\mathbf{\dot q}) \mathbf{\dot q} =  \mathbf{u}
\end{equation}
\begin{equation}
\label{eq:posconst}
\mathbf{q}_{min} \leq \mathbf{q}(t) \leq \mathbf{q}_{max}
\end{equation}
\begin{equation}
\label{eq:velconst}
\dot{\mathbf{q}}_{min} \leq \dot{\mathbf{q}}(t) \leq \dot{\mathbf{q}}_{max}
\end{equation}
\begin{equation}
\label{eq:actconst}
\mathbf{u}_{\min }\leq \mathbf{u}\left( t\right) \leq \mathbf{u}_{\max }
\end{equation}
\begin{equation}
\label{eq:initposconst}
\mathbf{q}(t_0) = \mathbf{q}_0,\ \ \ \mathbf{q}(t_f) = \mathbf{q}_f
\end{equation}
\begin{equation}
\label{eq:finvelconst}
\dot{\mathbf{q}}(t_0) = \dot{\mathbf{q}}_0,\ \ \ \dot{\mathbf{q}}(t_f) = \dot{\mathbf{q}}_f
\end{equation}
\end{subequations}

\autoref{eq:costfunctrajopt} shows the general cost function used with the running cost on time (with weight $w_t$) to find a minimum time solution and a quadratic cost on actuation $\mathbf{u}$ with weight $\mathbf{W_u} \in \mathbb{R}^{6+n \times 6+n}$. The terminal cost shown here is a quadratic cost on the final state of the system with weight $\mathbf{W_{x_f}} \in \mathbb{R}^{12+2n \times 12+2n}$. The cost on final velocity using the given terminal cost formulation can be used for the detumbling phase scenario in \autoref{subsec:detumbleresults}. \autoref{eq:trajoptdyneqn} - \autoref{eq:finvelconst} represent the constraints on the optimization problem. \comment[id=LS]{fixed reference (comment 6)}\autoref{eq:trajoptdyneqn} represents the dynamics constraints, \autoref{eq:posconst}, \autoref{eq:velconst}, and \autoref{eq:actconst} represent the position, velocity, and actuation limit constraints respectively, and \autoref{eq:initposconst} and \autoref{eq:finvelconst} represent the initial and final position and velocity constraints, if any.

The above-given non-linear optimization problem is transcribed into a non-linear program using direct transcription with Euler's integration scheme. The resulting non-linear problem is solved using the \textit{SNOPT (Sparse Nonlinear OPTimizer)} solver \citep{GilMS05} in Drake \citep{drake}. While higher order methods for trajectory optimization can be found in literature \citep{Betts2010, fahroo2002direct, shirazi2018spacecraft}, we reason that for the purpose of this work, the accuracy provided by the Euler integration scheme combined with $ 0.01 \leq \mathrm{d}t \leq 0.2$ is sufficient. The dynamics are satisfied at each knot point in the trajectory and the errors due to system identification of the model of the physical system are generally much larger than errors from the integration scheme in trajectory optimization.

The trajectory optimization method described here is used throughout both of the example scenarios in \autoref{text:results}. For the detumble scenario in \autoref{subsec:detumbleresults}, the aim is to bring the post-capture system to a stationary state with zero angular velocity i.e. detumbling. Thus, the terminal cost along with the terminal constraint on the velocity assists in the detumble trajectory optimization. For the experimental results in \autoref{subsec:elissaresults}, the position constraints along the trajectory were used to enforce a circular trajectory.


\subsection{Trajectory Stabilization using Time-Varying Linear Quadratic Regulators} 
\label{text:stabilization}

For trajectory stabilization using TVLQR, as the trajectory from trajectory optimization in the previous section leads to a stabilizable fixed point of the system, the infinite horizon stabilization of the trajectory can be guaranteed within the computed RoA. The optimal open-loop trajectories ($\mathbf{x}^*(t), \mathbf{u}^*(t)$) from \autoref{subsec:trajopt} cannot be directly run in a dynamics simulator or on the real-system due to integration errors, modeling inaccuracies, sensor inaccuracies and disturbances. Hence, a closed-loop control feedback policy is needed to stabilize the given trajectory. Here, TVLQR is employed to stabilize a trajectory leading to a fixed point $\mathbf{x}_f$, which is then stabilized using an infinite horizon LQR.
For the TVLQR synthesis, a time-varying linearization of the system along the trajectory is needed. For this, a moving reference frame is introduced along the trajectory:

\begin{equation}
\label{eq:err_coords}
\bar{\mathbf{x}}(t) = \mathbf{x}(t) -\mathbf{x^{*}}(t),\ \bar{\mathbf{u}}(t) = \mathbf{u}(t)-\mathbf{u^{*}}(t)
\end{equation}

The nonlinear dynamics are first linearized along the nominal trajectory $\mathbf{x}^{*}(t)$ and $\mathbf{u}^{*}(t)$ using first-order Taylor expansion yielding a linear time-varying system in the new coordinates described by \cite{Vyas2022a, Vyas2022}:

\begin{equation}
    \dot{\bar{\mathbf{x}}} = \mathbf{f(x,u)}- \mathbf{f(x^*,u^*)} \approx  \underbrace{\frac{\partial{\mathbf{f}(\mathbf{x}^{*},\mathbf{u}^{*}))}}{\partial{\mathbf{x}}}}_{:= \mathbf{A}(t)}\bar{\mathbf{x}}(t)+\underbrace{\frac{\partial{\mathbf{f}(\mathbf{x}^{*},\mathbf{u}^{*})}}{\partial{u}}}_{:=\mathbf{B}(t)}\bar{\mathbf{u}}(t) = \mathbf{A}(t) \bar{\mathbf{x}}(t) + \mathbf{B}(t) \bar{\mathbf{u}}(t)
\end{equation}

For this time-varying linear system, we can now define a quadratic cost function to drive the error to zero in the error coordinates thereby following the nominal trajectory. The cost function can be written as:

\begin{equation}
\label{eq:tvlqrCost}
\replaced[id=LS] 
{
J=\int_0^{t_f} ( \bar{\mathbf{x}}^T \mathbf{Q} \bar{\mathbf{x}} + \bar{\mathbf{u}}^T \mathbf{R} \bar{\mathbf{u}} ) dt + \bar{\mathbf{x}}_f^T \mathbf{Q}_f \bar{\mathbf{x}}_f
}
{
J=\int_0^{t_f} ( {\mathbf{x}}^T \mathbf{Q} {\mathbf{x}} + {\mathbf{u}}^T \mathbf{R} {\mathbf{u}} ) dt + {\mathbf{x}}_f^T \mathbf{Q}_f {\mathbf{x}}_f
}
\end{equation} 
\comment[id=LS]{comment 9} 

Where $\mathbf{Q} = \mathbf{Q}^T \succeq 0$ and $\mathbf{R} = \mathbf{R}^T \succ 0$ are positive semi-definite weight matrices that penalize deviations in state and actuation, respectively, along the trajectory and $\mathbf{Q}_f = \mathbf{Q}_f^T \succeq 0$ penalizes the error on the final state which is a fixed point. A closed loop optimal control policy for the linearized problem can be obtained by solving the differential Riccati Equation \citep{bertsekas2000dynamic}:

\begin{equation}
    \label{eq:diffRicatti}
    - \dot{\mathbf{S}}(t)=\mathbf{S}(t) \mathbf{A}(t) + \mathbf{A}^T(t) \mathbf{S}(t) - \mathbf{S}(t) \mathbf{B}(t) \mathbf{R}^{-1} \mathbf{B}^T(t) \mathbf{S}(t) + \mathbf{Q}
\end{equation}

\replaced[id=LS, comment=comment 10]
{
Integrating \ref{eq:diffRicatti} backwards with ${\bf S}(t_f) = {\bf Q}_f$ yields $\mathbf{S}(t)$, 
}
{
where $\mathbf{S}$ is 
}
the optimal cost-to-go. The feedback law can be written as:

\begin{equation}
    \label{eq:TVLQRFeedback}
    \mathbf{u}=\mathbf{u}^{*}(t)-\mathbf{K}(t)(\mathbf{x}(t)-\mathbf{x}^{*}(t))
\end{equation}

\added[id=LS, comment=comment 11]
{
Where 
\begin{equation}
    {\bf K}(t)   = - {\bf R}^{-1} {\bf B}(t)^T {\bf S}(t) 
\end{equation}
}

An advantage of utilizing TVLQR is, that the linear optimal cost-to-go $J^{*}(t) =\bar{\mathbf{x}}^T(t) \mathbf{S}(t) \bar{\mathbf{x}}(t)$ is a suitable candidate for a time varying Lyapunov function, which enables performing subsequent stability and robustness analysis of the closed loop dynamics.

\subsection{Region of Attraction}
\label{text:roa}
The region of attraction (RoA) $\mathcal{R}_a$ around a fixed point $\mathbf{x}^{*}$ of a nonlinear system $\dot{\mathbf{x}}=\mathbf{f}(\mathbf{x}(t))$ is the set of states that evolve to the fixed point as times approaches infinity. 
Oftentimes, there is no simple, closed-form description of the RoA for closed-loop systems. 
However, there exist multiple methods to estimate it, many of which build upon the notion of stability in the sense of Lyapunov. More specifically, the RoA can be estimated by considering sublevel sets of Lyapunov functions \citep{khalil2002}. 
Doing so, we define an inner approximation of $\mathcal{B} \subset \mathcal{R}_a$ by writing: 

\begin{equation}
    \label{eq:roa_basic}
    \mathcal{B}=\{ \mathbf{x} \vert V(\mathbf{x}) < \rho \}
\end{equation}

Assuming a Lyapunov function $V(\mathbf{x})$ is known, estimating the RoA is a matter of finding the greatest scalar 
\replaced[id=LS,comment=comment 14]
{
$\rho > 0$
}
{
$\rho$
}
, such that the Lyapunov conditions are satisfied, i.e. $V(\mathbf{x}) > 0$ and $\dot{V}(\mathbf{x}) = \nabla V \mathbf{f}(\mathbf{x}) \leq 0$ for $x \in \mathcal{B}$.
There exist multiple ways to estimate $\rho$.
One can use Sum of Squares programming (SoS) and convex optimization to reason about the problem algebraically and obtain a formal certificate for stability \citep{parrilo2000}.
However, formulating SOS programs can often be challenging, not only because of the background knowledge that is required but also because doing so requires expressing the closed-loop dynamics in a polynomial, non-rational form. 
Besides this optimization-based approach, there exist probabilistic methods for certifying stability in the sense of Lyapunov \citep{najafi2016}. 
For time-invariant systems, one can evaluate the right-hand side of $\mathbf{f}(\mathbf{x})$ for states $\tilde{\mathbf{x}}$ randomly sampled from the current estimate $\mathcal{B}$ and check whether the Lyapunov conditions are satisfied. 
If this is not the case, $\rho$ is reduced to $\rho = \mathbf{V}(\tilde{\mathbf{x}})$ and the next $\tilde{\mathbf{x}}$ is sampled from the shrunk estimate.
This process is then repeated until convergence and yields a probabilistic certificate of stability.

For time-varying dynamics $\dot{\mathbf{x}}=\mathbf{f}(\mathbf{x}(t),t)$, the notion of a funnel described by a sublevel set of a time-varying Lyapunov function can be seen as an extension of the RoA to time-varying systems~\citep{tedrake2010,tobenkin2011, majumdar2017}:

\begin{align}
\label{eq:funnel}
    \mathcal{B}(t) & =\left\{ \mathbf{x} \vert V(\mathbf{x},t) < \rho(t) \right\} \\
    \mathbf{x}(t) \in \mathcal{B}(t) & \implies \mathbf{x}(t_f) \in \mathcal{B}_f
\end{align}
where in contrast to \autoref{eq:roa_basic}, $\mathcal{B}(t)$, $V(\mathbf{x},t)$ and $\rho(t)$ are now time varying.
In \autoref{eq:funnel} $\mathcal{B}(t)$ describes a time-varying set of states leading to a set of goal states $\mathcal{B}_f$.
Similar as for the time invariant case, $\mathcal{B}(t)$ can be estimated by using SoS based \citep{majumdar2017, tedrake2010, moore2014}, or probabilistic \citep{reist2010, reist2016} methods.
Instead of evaluating the right-hand side as was done sampling based time invariant RoA estimation by \cite{najafi2016}, the approach presented by \cite{reist2010} relies on exhaustive simulations of the dynamics.

%% file: sections/roaest.tex
\begin{wrapfigure}{r}{0.4\linewidth}
    \centering
    \includegraphics[width=\linewidth]{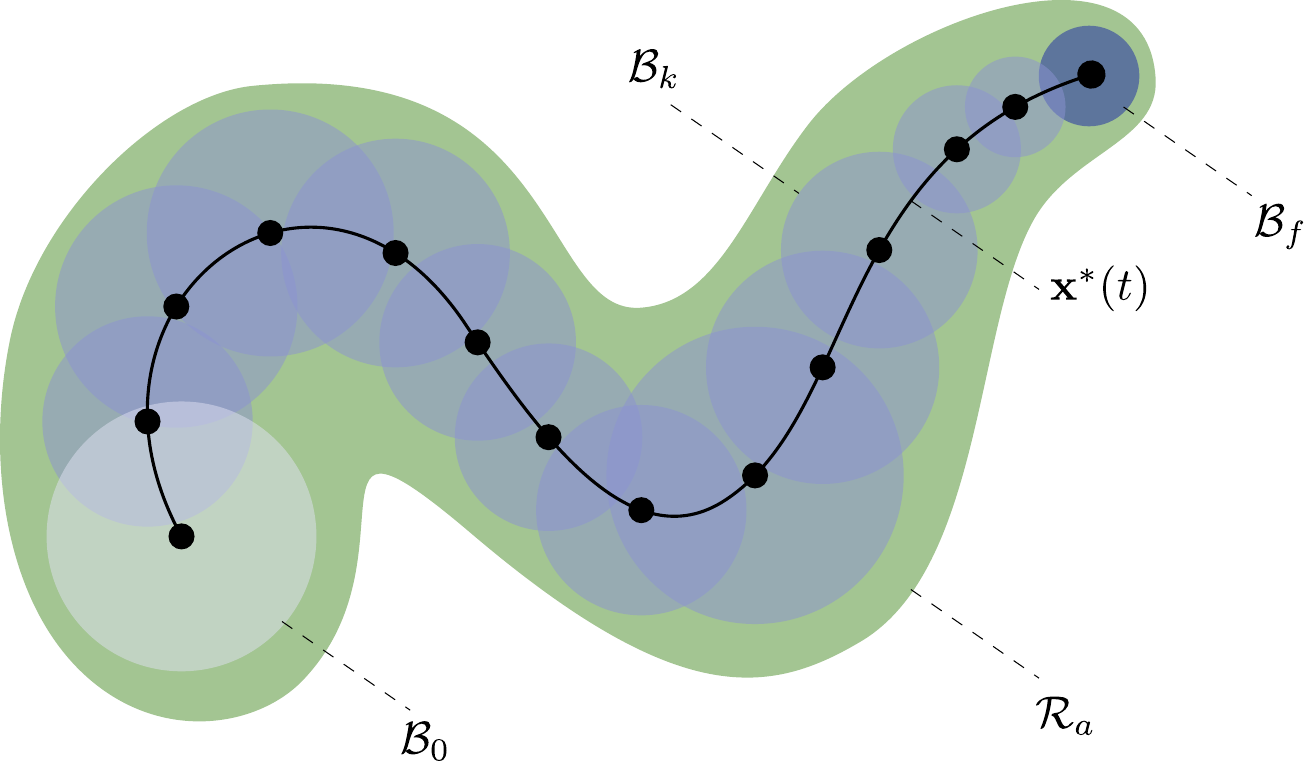}
    \caption[]{Time varying RoA estimate using a set of ellipses (blue). $\mathcal{B}_0$ and $\mathcal{B}_f$ are the inlet and the outlet of the funnel, respectively. $\mathcal{R}_a$ shows the actual, albeit unknown RoA  (green). The set of ellipses $\mathcal{B}_k$ is an approximation of the funnel $\mathcal{B}(t)$.}
    \label{fig:tvroa}
\end{wrapfigure}

\label{text:roa_est}
In order to estimate funnels according to the definition given in \autoref{text:roa} for closed loop dynamics that incorporate TVLQR feedback policies, a probabilistic approach based on \citep{reist2010} was chosen, mainly due to the greater flexibility when it comes to adding constraints or making changes to the model.
As stated in \autoref{text:stabilization}, considering the nonlinear closed loop dynamics $\mathbf{f}_{cl}$ of a multibody-system under TVLQR control, the quadratic cost-to-go $J^*$ can be used as a Lyapunov function, i.e. $V = J^* = \mathbf{x}^{T} \mathbf{S} \mathbf{x}$. 
Doing so yields the time-varying RoA (funnel) around $\mathbf{x}^{*}$:

\begin{equation}
\label{eq:funnel_cont}
    \mathcal{B}(t)=\left\{
    \replaced[id=LS]
    {\bar{\mathbf{x}}}
    {\mathbf{x}}
    \vert \bar{\mathbf{x}}^T(t) \mathbf{S}(t) \bar{\mathbf{x}}(t) < \rho(t) \right\}
\end{equation}
\comment[id=LS]{comment 16}

which, since $\mathbf{S}(t) \succ 0$, describes a time-varying hyperellipsoid. 
In practice, a discrete time version of \autoref{eq:funnel_cont} is used and $\rho$ is estimated around the collocation points of $\mathbf{x}^{*}$, which results in a set of hyperellipsoids as depicted in \autoref{fig:tvroa}.

The estimation process is based on simulating the closed loop dynamics $\mathbf{f}_{cl}$ forward in time for initial states randomly sampled from a continuously shrinking estimate of the inlet of the estimated funnel $\mathcal{B}_0 = \mathcal{B}(t=0)$. 
More specifically, initial states that after integration of $\mathbf{f}_{cl}$ do not lead to a final state with a linear cost-to-go $J^{*}(t_{\text{f}}) < \rho_f$ or violate constraints, cause shrinking of the hyperellipsoids that make up the funnel, including its inlet $\mathcal{B}_0$. 
Using this methodology, funnels can only be shrunk, never expanded.
Pseudocode for the algorithm used in this work can be found in \autoref{alg:tvlqr}.
\begin{algorithm}
    \caption{Probabilistic RoA Estimation for TVLQR} 
	\begin{algorithmic}[1]
	    \State $\mathbf{x}^{*}(t)$, $\mathbf{u}^{*}(t)$ $\leftarrow$ nominal trajectory 
	    \State $\mathbf{A}(t)$, $\mathbf{B}(t)$ $\leftarrow$ linearized dynamics along $\mathbf{x}^{*}(t)$, $\mathbf{u}^{*}(t)$ 
	    \State $\mathbf{S}(t)$, $\mathbf{K}(t)$ $\leftarrow$ Finite Horizon LQR for $\mathbf{A}(t)$, $\mathbf{B}(t)$ 
	    \State $n$ $\leftarrow$ Nr. of Simulations
	    \State $m$ $\leftarrow$ Nr. of Evaluation Steps 
	    \State $\rho_{0,k}=\infty,k \in \{ 0,\dots,n-1 \}$
	    \State $\rho_{0,f}=\rho_n=\bar{\mathbf{x}}^T_{\text{f},\max} \mathbf{S}_{\text{f}} \bar{\mathbf{x}}_{\text{f},\max}$ \Comment{Definition of goal region}
        \For{$j=1,2,\dots,n$}
            \State shrink=\textbf{false}
            \State $\mathbf{x}_0$ $\leftarrow$ random initial state from $\{ \mathbf{x} \vert \bar{\mathbf{x}}^T_{0} S_{\text{0}} \bar{\mathbf{x}}_{0} < \rho_{j-1,0} \}$
            \State ${J}^{*}_{j,0}=\bar{\mathbf{x}}^T_{0} S_{0} \bar{\mathbf{x}}_{0}$
            \For{$k=1,2,\dots,m$}
                \State Integrate $\mathbf{f}_{cl}$ until $t=t_k$
                \If{Constraint Violation}
                    \State shrink=\textbf{true}
                    \State \textbf{break} 
                \Else
                    \State ${J}^{*}_{j,k}=\bar{\mathbf{x}}^T_{k} S_{k} \bar{\mathbf{x}}_{k}$ \Comment{compute \emph{cost-to-go}}
                    \If{${J}^{*}_{j,k} > \rho_{j-1,k}$}
                        \State shrink=\textbf{true}
                        \State \textbf{break} 
                    \EndIf
                \EndIf
            
            \EndFor
        \If{shrink == \textbf{true}}
            \State $\rho_{j,\kappa}={J}^{*}_{j,\kappa}, \kappa \in \{0,1,\dots,k-1\}$ \Comment{update estimates}
        \EndIf
        \EndFor
	\end{algorithmic} 
	    \label{alg:tvlqr}
\end{algorithm}

We initialize our estimate of $\rho$ with $\rho=\infty$ along the trajectory, except for $\rho_f$, associated to the outlet $\mathcal{B}_f = \mathcal{B}(t=t_f)$ (\autoref{alg:tvlqr}, lines 6 and 7). 
The value of $\rho_f$ remains fixed throughout the estimation process and can be defined by considering the cost-to-go $J^{*}_{\text{max}}$ of a maximum allowed deviation to the nominal final state $\bar{\mathbf{x}}^T_{\text{max}}(t_{\text{f}})$ as described in~\cite{cory2010}:

\begin{align}
    \label{eq:rhoFinFF}
    \rho_{f} =\rho(t_{f})& =J^{*}_{\text{max}}(t_{f}) =\bar{\mathbf{x}}^T_{\text{max}}(t_{f}) \mathbf{S}(t_{f}) \bar{\mathbf{x}}_{\text{max}}(t_{f}) 
\end{align}
where $\bar{\mathbf{x}}^T_{\text{max}}$ is the user-defined, tolerable maximum deviation from the final state.
In order to ensure infinite horizon stability, $\rho_{\text{f}}$ must be chosen, such that the outlet $\mathcal{B}_f \subset \mathcal{B}_{\infty}$, where $\mathcal{B}_{\infty}$ is the RoA associated to the closed loop dynamics (infinite horizon LQR) around a fixed point at the end of $\mathbf{x}^{*}$.
After initializing $\rho_0$ and $\rho_f$, a random initial condition is sampled from $\mathcal{B}_0$ (\autoref{alg:tvlqr}, line 10) using a similar method as was used in~\citep{maywald2022}.
We found simple rejection sampling to be infeasible for sampling from a constantly shrinking $\mathcal{B}_0$ in a potentially high dimensional state space.
The reason is the decline in acceptance rate of samples with increasing dimensionality of the sampling space \citep{krauth2006}.
Hence, direct sampling was employed, which is more complicated, but also more efficient and even required for systems with higher dimensional state spaces, such as the one that will be discussed in \autoref{subsec:detumbleresults}.
States are sampled from the $n$ dimensional unit sphere using the \textit{direct-sphere} algorithm found in~\cite{krauth2006} and then mapped to $\mathcal{B}_0$, the hyperellipsoid that describes the inlet of $\mathcal{B}$ in \autoref{eq:funnel_cont}:
\begin{wrapfigure}{R}{0.6\linewidth}
    \centering
    \includegraphics[width=\linewidth]{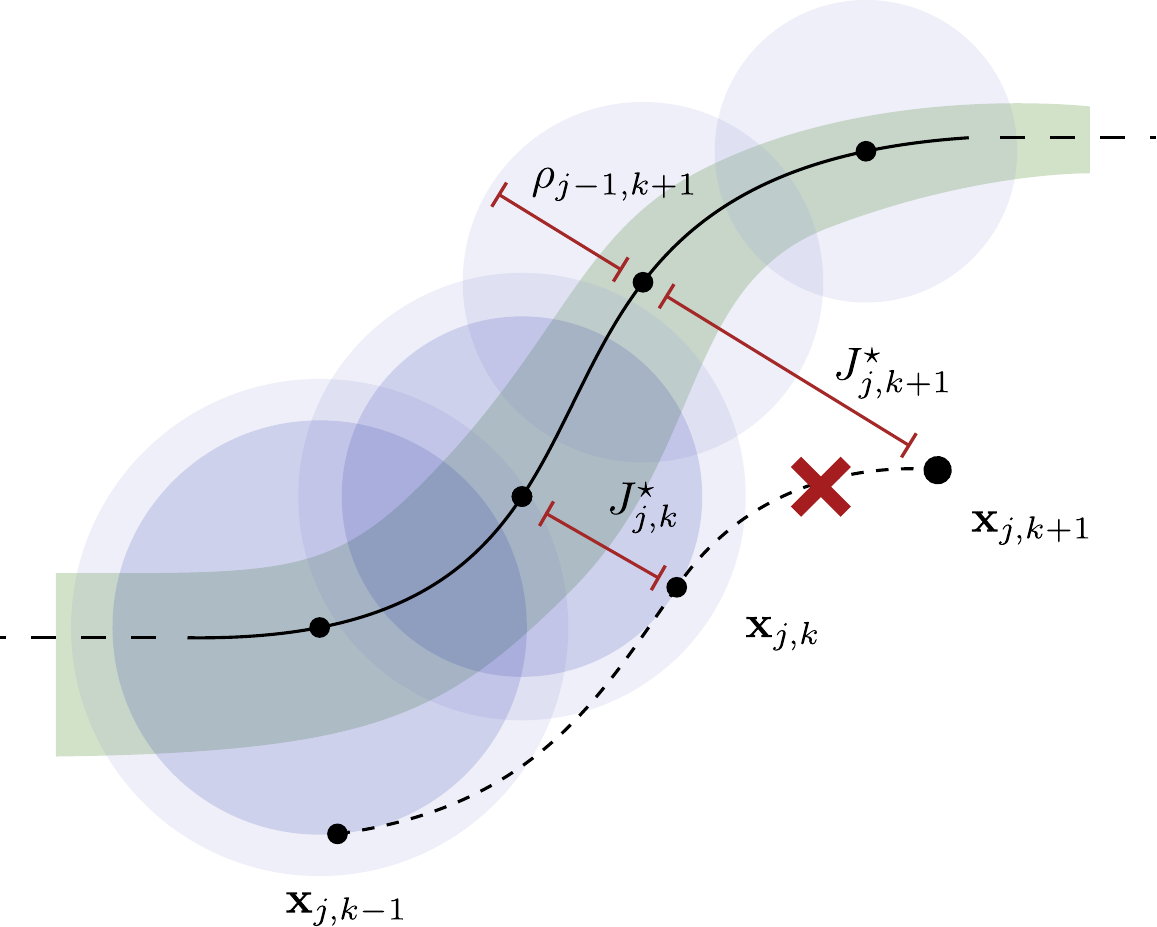}
    \caption{Estimation process for a hypothetical, two-dimensional system. Whenever the dynamics lead outside the previously estimated funnel (light blue ellipses), or a constraint has been violated (red "X") all preceding estimates have to be shrunk (darker blue ellipses). The green area shows the real funnel, that is iteratively being approximated by the set of ellipses.}
    \label{fig:shrinking}
\end{wrapfigure}
\begin{equation}
    \label{eq:first_ell}
    \mathcal{B}_0 = \left\{ \mathbf{x} \vert \bar{\mathbf{x}}_0^T \mathbf{S}_0 \bar{\mathbf{x}}_0 \leq \rho_0 \right\}.
\end{equation}

Consider the sphere $\mathcal{S}_n = \left\{ \mathbf{y} \vert \mathbf{y}^T \mathbf{y} \leq 1 \right\}$ and a linear mapping  $\mathbf{T} \in \mathbb{R}^{n \times n}$, so that $\mathbf{y}=\mathbf{T}\mathbf{x}$. This allows to express $\mathcal{S}$ in terms of $\mathbf{x}$:
\begin{equation}
\label{eq:nsphere}
    \mathcal{S}_n = \left\{ \mathbf{x} \vert \mathbf{x}^T \mathbf{T}^T \mathbf{T} \mathbf{x} \leq 1 \right\}
\end{equation}
Comparing \autoref{eq:first_ell} with \autoref{eq:nsphere}, we now seek to find $\mathbf{T}$, such that $\mathbf{T}^{T} \mathbf{T} = \rho_0^{-1} \mathbf{S}_0$.
A diagonalization of $\mathbf{T}^{T} \mathbf{T}$ yields:
\begin{equation}
    \mathbf{T}^{T} \mathbf{T} = \mathbf{W}^{T} \bm{\Lambda} \mathbf{W}
\end{equation}
where $\mathbf{W}$ and $\bm{\Lambda}$ are matrices of eigenvectors and -values of $\rho_0^{-1} \mathbf{S}_0$, respectively.
With $\mathbf{T}=\bm{\Lambda}^{\frac{1}{2}} \mathbf{W}$, the mapping from a random $\mathbf{y} \in \mathcal{S}_n$ to a random $\mathbf{x}_0 \in \mathcal{B}_0$ is:
\begin{equation}
    \mathbf{x}_0 = \mathbf{x}^*_0 + \left( \Lambda^{\frac{1}{2}} \mathbf{W} \right)^{-1} \mathbf{y}
\end{equation}
In the following, we describe the shrinking operation (visualized in \autoref{fig:shrinking}).
For this, let $m$ and $n$ denote the number of simulations made during the estimation process and the number of knot points of $\mathbf{x}^{*}$, respectively.
Furthermore, $j \in [0, \dots m]$ and $k \in [0, \dots n]$, denote the index associated with the current simulation and knot point, respectively.
After sampling an initial state $\mathbf{x}_{j,0}$, the closed loop dynamics $\mathbf{f}_{cl}$ are integrated piece-wise in between knot points that are evenly spaced wrt. time.
At each knot-point, where $t_k= k \Delta t $, the linear optimal cost-to-go $J^{*}_{j,k}$ of the current state $\mathbf{x}_{j,k}$ is compared to that of the associated value of $\rho_{j-1,k}$ made in the previous, that is $(j-1)$th, simulation. 
If $\mathbf{x}_{j,k}$ is outside the existing funnel estimate, that is, $J^{*}_{j,k} > \rho_{j-1,k}$, all estimates of $\rho$ made for previous knot-points within preceding simulations ($\rho_{j-1,k-1},\rho_{j-1,k-2},\dots,\rho_{j-1,0}$) have to be adjusted to the optimal cost-to-go values estimated during this simulation, i.e. $\rho_{j,k} = J^{\star}_{j,\kappa}$ for $\kappa \in \{ 0, \dots, k \}$ (\autoref{alg:tvlqr}, line 18). 
Similarly, if during the integration of $\mathbf{f}_{cl}$ a constraint is violated, we resize the funnel (\autoref{alg:tvlqr}, line 14).
Thus, the shrinking policy can be written as:
\begin{equation}
    \label{eq:shrinkage}
    \rho_{j,\kappa} = \left\{\begin{array}{@{}ll@{}}
    \multirow{2}{*}{$J^{\star}_{j,\kappa}$   \ \ \ \ \ for  $\kappa \in \{ 0, \dots, k \},$} &  \text{if constraint violated  (alg 1., line 15)} \\
    & \text{or } J^{*}_{j,k+1} > \rho_{j-1,k+1} \text{\ (alg 1., line 21)}     \vspace{0.2cm}\\
    \rho_{j-1,\kappa}  \text{\ \ for }   \kappa \in \{ 0, \dots, n-1 \}, & \text{else }
    \end{array} \right.
\end{equation}
After aborting or finishing a simulation successfully, we continue by sampling the next initial state and start over until convergence can be seen, or $j=m$.

%% file: sections/results.tex
\label{text:results}
Two scenarios have been conceived to investigate the applicability of TVLQR and RoA estimation for space robotics. First, using the RoA estimation methodology introduced in \autoref{text:roa}, the funnel of a multi-body detumbling scenario has been computed in simulation. Secondly, in order to demonstrate TVLQR on a physical free-floating system, experiments have been conducted onboard the Experimental Lab for Proximity Operations and Space Situational Awareness (\textsc{ELISSA}), a free-floating laboratory.

\subsection{Scenario I: Simulated Multibody ADR Detumbling Maneuver}
\label{subsec:detumbleresults}

\begin{wrapfigure}{r}{0.4\linewidth}
    \centering
    \includegraphics[width=\linewidth]{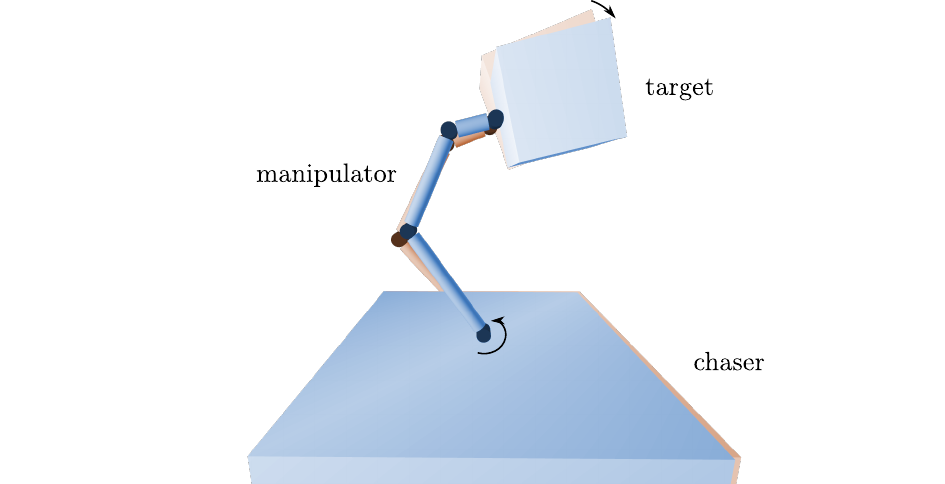}
    \caption{Chaser (partially visible) and target satellite are connected by a 3R manipulator. In the beginning (orange) the chaser is rotating. After executing the detumbling controller, there is no relative velocity of both spacecraft (blue).}
    \label{fig:scenario1}
\end{wrapfigure}

A post-capture detumbling scenario has been conceived, in which a chaser satellite with a 3-DoF robot arm has already docked to a target debris object (as shown in \autoref{fig:scenario1}), that is still rotating with an angular velocity of $\omega_0 = \SI{5}{\degree\per\second}$. Assuming perfect synchronization of the grasping mechanism and the capture point, it can be assumed that there were no contact forces involved during the capture i.e. ideal capture. This allows the post-capture capture velocities to be determined using the Generalized Jacobian Matrix \citep{Umetani1989} when assuming the center of mass of the target is the end-effector frame for the combined post-capture system. The chaser and target were modeled as cubes with side lengths of $l_c=\SI{2}{\meter}$ and $l_t=\SI{0.6}{\meter}$, and masses of $\SI{100}{\kilo\gram}$ and $\SI{50}{\kilo\gram}$, respectively and with a homogenous mass distribution. The manipulator's link masses are $m_1=\SI{10}{\kilo\gram}$, $m_2=\SI{8}{\kilo\gram}$ and $m_3=\SI{4}{\kilo\gram}$ and their respective lengths are $l_1=\SI{0.9}{\meter}$, $l_2=\SI{0.7}{\meter}$ and $l_3=\SI{0.3}{\meter}$.
The dynamics of this system, described by \autoref{eq:dynamicseqn} were linearized according to~\cite{Vyas2022a}, so that the state vector $\mathbf{x}$ can be written as:

\begin{equation}
    \mathbf{x}=
    \begin{bmatrix}
    \mathbf{\hat q}_B & \mathbf{p}_B & q_1 & q_2 & q_3 & \bm{\omega}_B & \mathbf{v}_B & \dot{q_1} & \dot{q_2} & \dot{q_3}
    \end{bmatrix}^T
\end{equation}
where $\mathbf{\hat q}_B$ denotes the vector part of a quaternion, $\mathbf{p}_B \in \mathbb{R}^3$ is the position of the floating base and $q_1$, $q_2$ and $q_3$ specify the manipulator's configuration. 
Angular and translational velocities are given by $\bm{\omega}_B \in \mathbb{R}^3$ and $\mathbf{v}_B \in \mathbb{R}^3$, respectively. 
The manipulators joint velocities are written as $\dot{q_1}$, $\dot{q_2}$ and $\dot{q_3}$.
The complete input vector $\mathbf{u}$ is:
    
\begin{equation}
\mathbf{u}=
\begin{bmatrix}
\tau_x & \tau_y & \tau_z & f_x & f_y & f_z & \tau_1 & \tau_2 & \tau_3 
\end{bmatrix}^T
\end{equation}
where $\tau_x$,$\tau_y$, $\tau_z$, $\tau_1$,$\tau_2$ and $\tau_3$ are torques applied to the base and manipulator, respectively.
In addition, the base is assumed to be equipped with thrusters, such that forces $f_x$,$f_y$ and $f_z$ can be applied.
Actuation limits were set to $\pm 50 \si{\newton\meter}$ for the base and manipulator torques and $\pm 10 \si{\newton}$ for the base forces. A detumble trajectory was found using the above-given model and actuation using the transcription method in \autoref{subsec:trajopt}. This trajectory was stabilized using the method given in \autoref{text:stabilization} and executed on the Drake dynamics simulator \citep{drake}.

For RoA estimation, a generalized fuel constraint has been implemented, that does not make any specific assumptions of the actuators used. Instead, the idea of generalized forces applied at the center of mass of the base for the satellite and at the joints of the robot arm was employed. The fuel spent at time t is calculated as the integral of the sum of all elements of $\mathbf{u}(t)$:

\begin{equation}
    \label{eq:fuel}
    F(t) = \int_{\tau=0}^{t} \sum_{k=0}^{9} \vert u_k(t) \vert d \tau
\end{equation}
 
We calculate $F_0$, the fuel usage of the nominal trajectory by integrating \autoref{eq:fuel} over the duration of the entire nominal trajectory (with $\mathbf{u} = \mathbf{u}^{*}$). Furthermore, we define the maximal fuel usage for the entire maneuver as:

\begin{equation}
    \label{eq:fuel_constr}
    F_{\max} = \left( 1+\alpha \right)F_0
\end{equation}
where $\alpha$ is a parameter that relates the nominal amount of fuel to the share that TVLQR is allowed to use for stabilization.
While carrying out the RoA estimation, \autoref{eq:fuel} is evaluated and compared to $F_{\max}$. 
If $F(t) > F_{\max}$, the simulation loop is aborted (\autoref{alg:tvlqr}, line 14).

\begin{wrapfigure}{r}{0.75\linewidth}
    \centering
    \includegraphics[width=\linewidth]{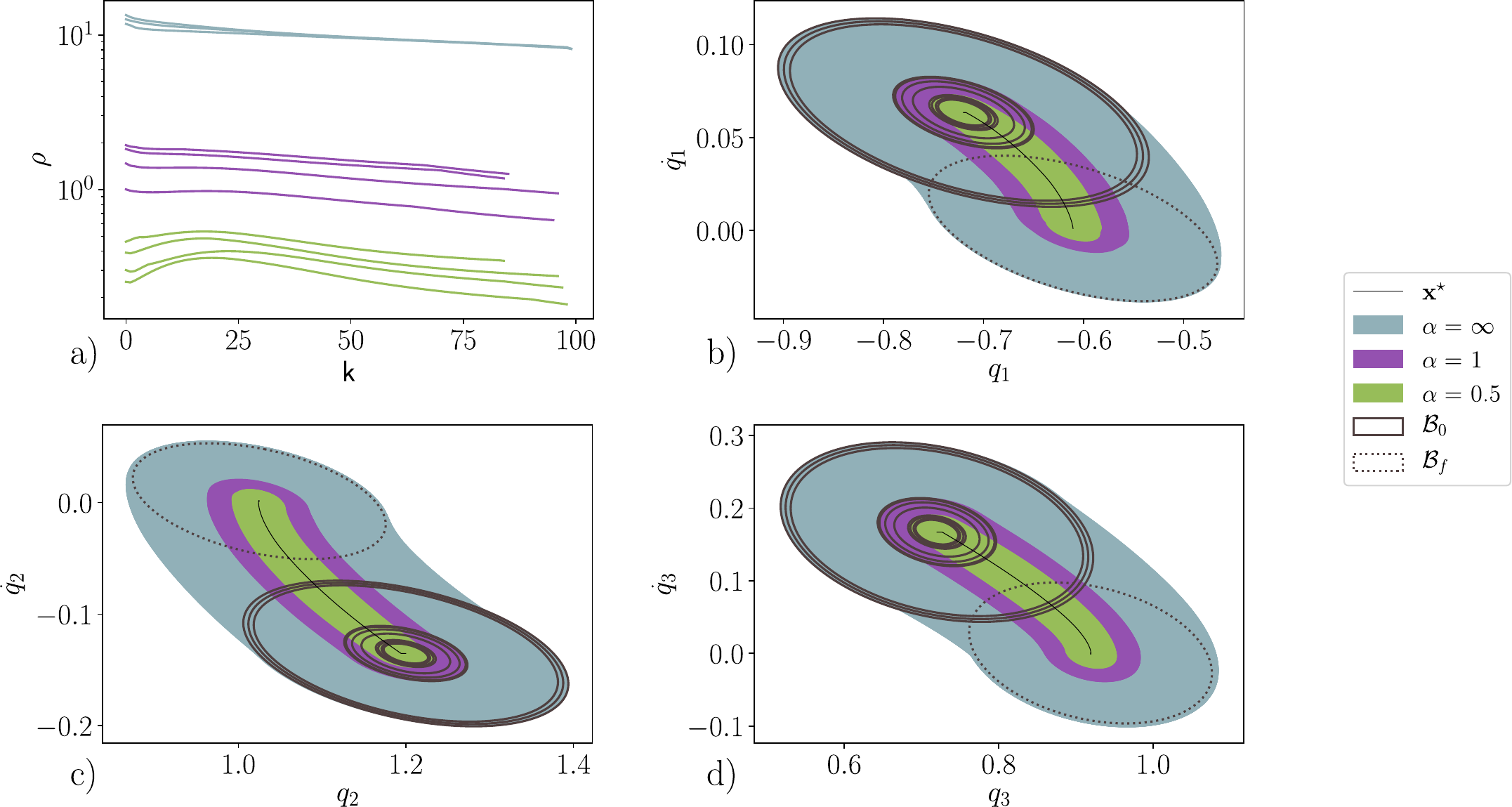}
    \caption{a) Evolution of $\rho$ for different fuel limit settings and over multiple estimation processes. b) to d) projections of the estimated RoA into different subspaces.}
    \label{fig:funnels_alpha}
\end{wrapfigure}

First, the RoA was calculated for $\alpha = \infty$, essentially bypassing the fuel limit. 
Furthermore, values of $\alpha=1$ and $\alpha=\frac{1}{2}$ were used.
For each $\alpha$, 5 RoA estimations were conducted with $n=1000$ simulations each, and $m=100$ knot points.
\autoref{fig:funnels_alpha} a) shows the final values of $\rho$ for three different fuel limit settings. 
One can see, that while estimates for equal values of $\alpha$ lie close together, there is a discrepancy when seeding the random number generator differently.
This is caused by the probabilistic nature of the estimation process in combination with the relatively large state space and the limited number of simulations that could be made within a reasonable time of $\approx 8$ hours on a recent laptop (Intel i9, 32GB RAM).
Furthermore, one can see that some of the lines stop before $k=100$ because some RoA estimates associated with knot points at the end of the trajectory did not undergo any shrinking operations, such that $\rho$ remained at its initialization value, i.e. $\rho = \infty$.
A more intuitive view of the estimated funnel is given in  \autoref{fig:funnels_alpha} b) to \autoref{fig:funnels_alpha} d), which show projections of the funnel into various 2d slices of the state space under the assumption that all remaining states are held nominal. 
Here, funnels associated with the different values of $\alpha$ are shown in different colors. 
The inlets and outlets are signified by solid and dashed ellipses, respectively.
For example, \autoref{fig:funnels_alpha} b) shows, that if $\alpha=\infty$, all states contained within $\mathcal{B}_{0,\infty}$ (the outermost solid ellipse) can be brought to a final state that has a linear optimal cost-to-go $J^*_f$ that is smaller than that of the states that lie on the border of $\mathcal{B}_{f}$.
Again, the fact that estimates do not completely converge to the same result is visible in the projections.

\subsection{Scenario II: Experiments onboard Elissa}
\label{subsec:elissaresults}
\begin{figure}
\centering
\includegraphics[width=0.75\linewidth]{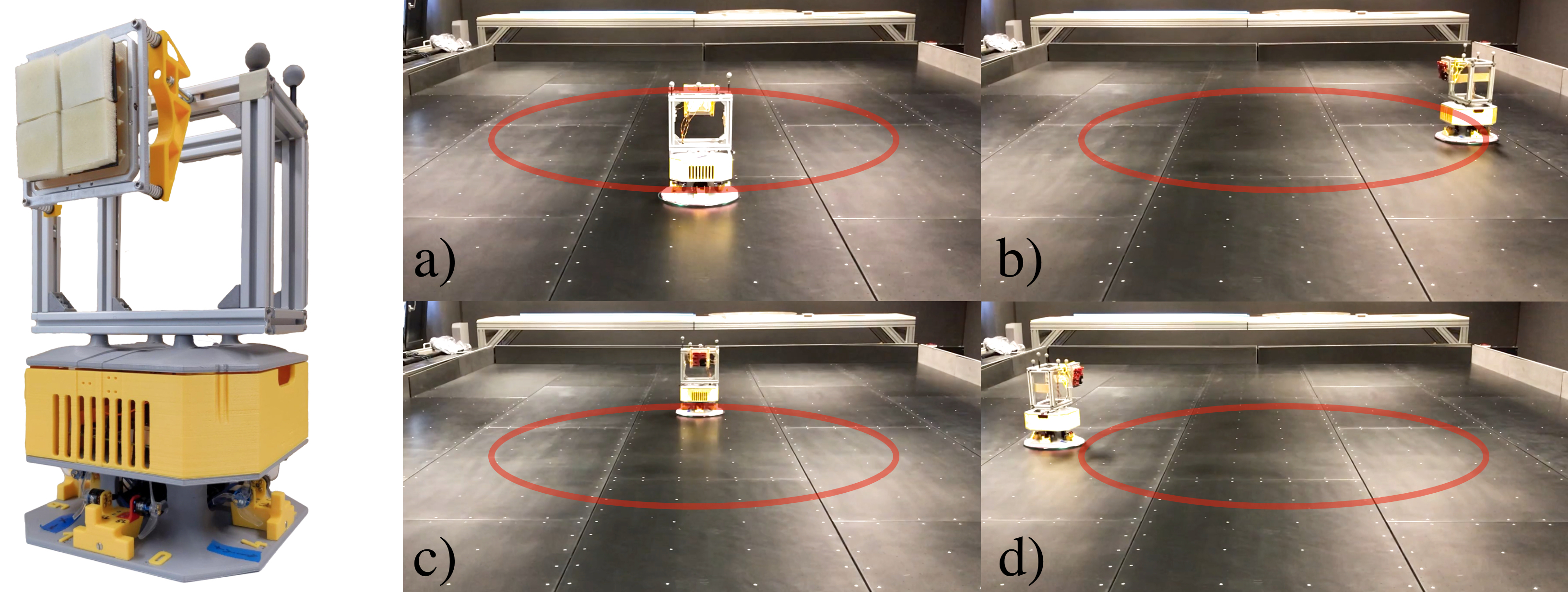}
\caption[Photographs of \emph{Hamilcar} moving along the nominal trajectory]{Photographs of \emph{Hamilcar} (left) moving along the nominal trajectory (right).}
\label{fig:elissaSeq}
\end{figure}

To demonstrate the capability of TVLQR and probabilistic RoA estimation for real-world free-floating systems, further lab experiments have been conducted onboard the ELISSA free-floating laboratory. 
ELISSA consists of several free-flyers (\autoref{fig:elissaSeq}) that float on an active air-bearing table. 
\begin{wrapfigure}{r}{0.5\linewidth}
\centering
\includegraphics[width=\linewidth]{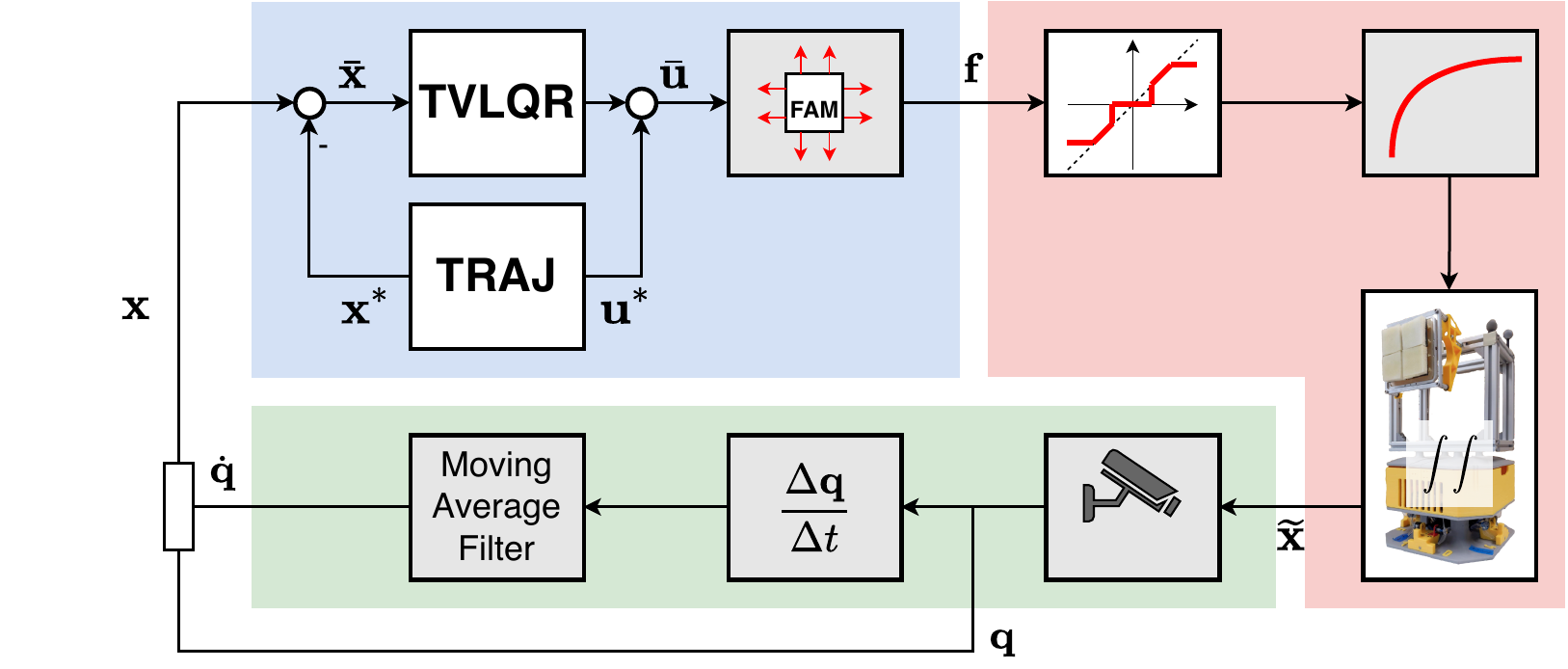}
\caption[Results]{Block diagram of ELISSA. Components with a gray background were not taken into account by RoA estimation. Blue color denotes control-related blocks, red refers to the plant, and green to state estimation.}
\label{fig:elissa}
\end{wrapfigure}
Said table has numerous small vents through which air can flow, thereby creating a thin cushion of air between its surface and objects on it. 
A block diagram of the experiment setup can be found in \autoref{fig:elissa}, which shows components that were not taken into account in the RoA estimation with a gray background.
The desired input $\mathbf{u}$ is calculated according to \autoref{eq:TVLQRFeedback} and fed into a force allocation block (FAM), that calculates the actuator forces $\mathbf{f}$ required to realize $u$ according to \citep{zappulla2017, jonckers2022}.
The required forces are then translated into PWM signals and sent to the motor drivers and actuators, subsequently. 
The actuators exhibit a deadband characteristic and take a while to deliver the demanded thrust.
An optical tracking system was utilized to track the freeflyers attitude $\theta$ and position $p_x, p_y$. 
Angular and linear velocities were calculated by numerical differentiation and then smoothened by a moving average filter before being fed back into the controller.
Within this work, the freeflyer was used in a hardware-in-the-loop configuration, meaning that the controller was executed on a lab PC. ROS Melodic was used for message passing and the control frequency was $\approx 30 Hz$.
Each freeflyer is modeled as a single rigid body ($m=4.26kg$, $I_{zz} = 0.064 kgm^{2}$) and uses eight small propeller-based thrusters for actuation. The propeller's dynamics are not taken into account during trajectory optimization and TVLQR synthesis. 
The state $\mathbf{x} \in \mathbb{R}^6$ of the 3-DoF system can be written as:
\begin{equation}
\mathbf{x} = \begin{bmatrix} \mathbf{q} & \dot{\mathbf{q}} \end{bmatrix}^T  = \begin{bmatrix} p_x & p_y & \theta_z & v_x & v_y & \omega_z \end{bmatrix}^T     
\end{equation}
and the input $\mathbf{u} \in \mathbb{R}^3$ can be written as $\mathbf{u} = \begin{bmatrix} f_x & f_y & \tau_z  \end{bmatrix}^T $.

Based on \autoref{eq:trajopt}, direct collocation-based trajectory optimization was used to compute a feasible nominal trajectory assuming linear dynamics. 
The transcribed problem is given in \autoref{eq:elissaTranscribed}

\begin{subequations}
\label{eq:elissaTranscribed}
    \begin{alignat}{8}
    \min _{\mathbf{x}_{k}, \mathbf{u}_{k}} \quad &\sum_{k=0}^{100} \Delta t_{k}\left( W_{t}+ \mathbf{u}_{k}^{T} \mathbf{W}_{u} \mathbf{u}_{k}\right) \\
    \textrm{s.t.} \quad & \mathbf{q}_{k+1}= \mathbf{q}_{k}+\Delta t_k \dot{\mathbf{q}}_{k} \label{eq:qint} \\
                        & \dot{\mathbf{q}}_{k+1}= \dot{\mathbf{q}}_{k}+\Delta t_k \ddot{\mathbf{q}}_k \label{eq:dqint} \\
                        & \ddot{\mathbf{q}}_{k}-\mathbf{M}^{-1} \mathbf{u}_{k}=0 \label{eq:actuation} \\
                        & \Delta t_{k+1}=\Delta t_{k} \label{eq:timestep} \\
                        & -\mathbf{u}_{\max } \leq \mathbf{u}_k \leq \mathbf{u}_{\max } \label{eq:limits}\\
                        & \mathbf{q}_{30} = \begin{bmatrix} 3 & 1 & \frac{\pi}{2}\end{bmatrix}^{T} \label{eq:q30} \\
                        & \mathbf{q}_{50} = \begin{bmatrix} 4 & 2 & \pi \end{bmatrix}^{T}\\
                        & \mathbf{q}_{70} = \begin{bmatrix} 3 & 3 & \frac{3 \pi}{2}\end{bmatrix}^{T} \label{eq:q70} \\
    \end{alignat}
\end{subequations}
Where $\Delta t_{k}$ is the integration step size, and the cost consists of $W_t$ penalizing the overall duration of trajectories and a quadratic cost on $\mathbf{u}$ with a weighing defined by $\mathbf{W}_u \in \mathbb{R}^{3 \times 3}$. 
A simple box quadrature was used as an integration scheme to model the double integrator dynamics (\autoref{eq:qint} and \autoref{eq:dqint}):
\begin{equation}
    \mathbf{q}_{k+1}-\mathbf{q}_{k}= \int_{t_k}^{t_{k+1}} \dot{\mathbf{q}}(t) dt \approx \left(t_{k+1}-t_k \right) \dot{\mathbf{q}}(t_k) = \Delta t_k \dot{\mathbf{q}}_k
\end{equation}
Furthermore, \autoref{eq:actuation} describes, how $\mathbf{u}$ affects the freeflyer, \ref{eq:timestep} forces equal time steps, and \ref{eq:limits} describes the actuation limits (not taking into account the deadband).
Intermediate waypoints (\autoref{eq:q30} to \ref{eq:q70}) were set to find a nearly circular trajectory.

A TVLQR policy was obtained according to the methodology described in \autoref{text:stabilization}. 
The weighing matrices have been set to $\mathbf{Q} = \text{diag}(50, 50, 0.01, 50, 50, 0.001)$ and $\mathbf{R} = \text{diag}(1,1,10)$. 
\autoref{eq:diffRicatti} was initialized with $\mathbf{S}\left( t_f \right) = \mathbf{Q}_f = \mathbf{S}_\infty$ the solution of the infinite Horizon LQR controller around the first and final state of the trajectory. 

\begin{figure}
\centering
\includegraphics[width=0.95\linewidth]{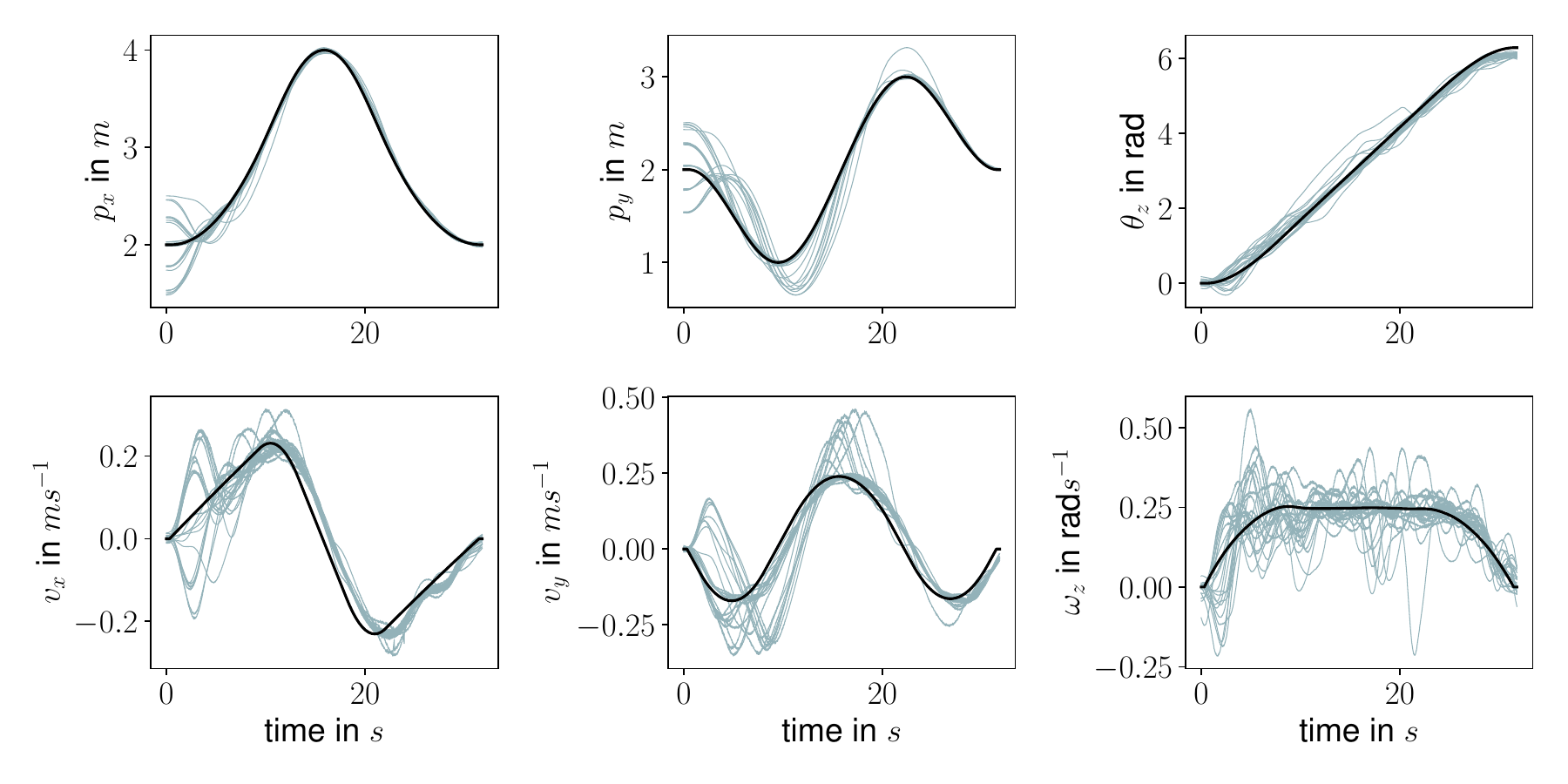}
\caption[Results]{Measured trajectories (blue) converging to the nominal trajectory (black).}
\label{fig:results}
\end{figure}

First, to verify and demonstrate the capabilities of TVLQR for free-floating systems, the control policy was executed for a set of off-nominal initial positions around $\mathbf{x}_0$. 
More specifically, initial positions were sampled from a $5 \times 5$, equally spaced, quadratic grid with a side length of $1m$. 
The resulting trajectories of the freeflyer as measured by the tracking system are shown in \autoref{fig:results}. 
For every initial condition (blue lines), convergence to the nominal trajectory (black line) can be observed, after the positional offset was corrected. 

\begin{figure}
\centering
\includegraphics[width=\linewidth]{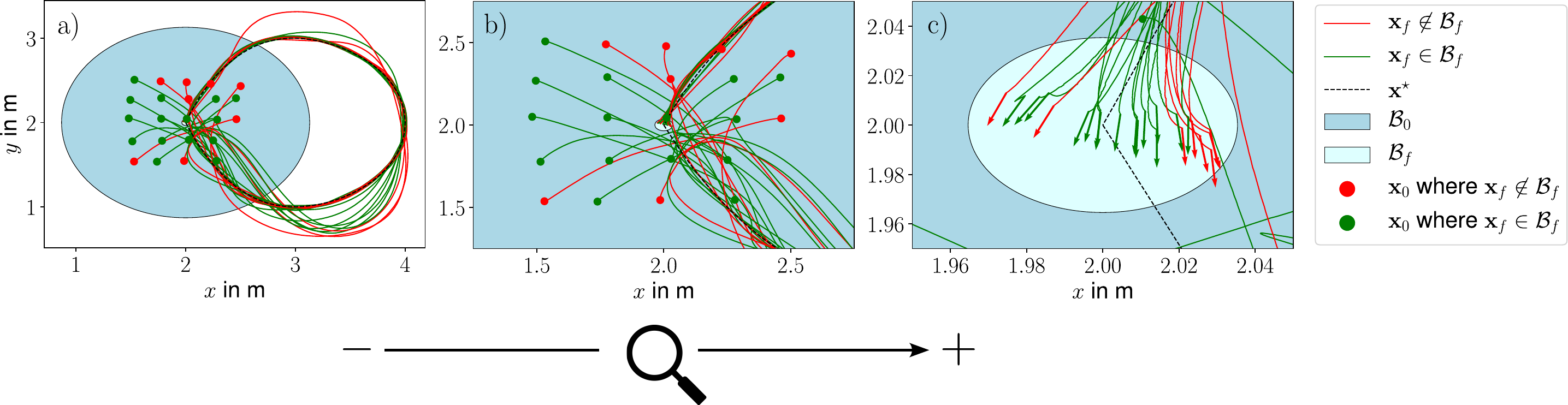}
\caption[Results]{Trajectories in the $x$ vs $y$ plane. Green and red indicate that the final state was in or outside the estimated RoA, respectively. Projections of the In- and Outlet of $\mathcal{B}$ are shown as dark and light-blue ellipses, respectively. Quivers at the end of the trajectory show the magnitude and direction of the terminal velocity.}
\label{fig:roa_elissa}
\end{figure}

In addition to demonstrating, that TVLQR could indeed stabilize the trajectories, a RoA estimation was carried out according to the method described in \autoref{text:roa}. 
The results are shown in \autoref{fig:roa_elissa}. 
The light- and dark-blue ellipses are projections of the funnel in- and outlet for the given nominal trajectory (black dashed line) and controller.  
Each circle represents one of the 23\footnote{Shortly before starting the 24th run, the freeflyers battery was drained.} initial states $\mathbf{x}_0$, all sampled from within the estimated inlet of the funnel $\mathcal{B}_0$. 
The solid lines are the measured trajectories of the freeflyer. 
Green and red color gives information on whether a given initial state/trajectory leads towards the predefined goal region $\mathcal{B}_f$ or not. 
At first glance, it appears as if every measured trajectory leads to a state within $\mathcal{B}_f$.
However, recall that the ellipses are merely a projection of $\mathcal{B}_f$ under the assumption that $v_x = v_y = 0$. 
Since freeflyers had nonzero terminal velocity (as indicated by the quivers in \autoref{fig:roa_elissa} b)), this assumption breaks. 
One can see, that trials for which $\mathbf{x}_f \not\in \mathcal{B}_f$ usually started at the outer domain of the sampled space of initial conditions and ended close to the bounds of the projected $\mathcal{B}_f$, which makes sense, since the magnitude of the terminal velocity is relatively similar across trials and has a larger influence on the cost if the positional deviation is larger.

%% file: sections/discussion.tex
\label{text:discussion}
As stated in \autoref{text:roa}, probabilistic methods such as the one conceived in \autoref{text:roa_est} have the big advantage of being applicable to a large range of systems.
As long as the system can be simulated, it does not matter what kind of dynamics and constraints it exhibits.
This is not the case for SoS based methods, where every dynamics constraint has to be carefully transcribed into a constraint within an optimization problem. 
Furthermore, the fact that only non-rational polynomial functions can be used, limits the range of applicable systems or introduces inaccuracies, if one decides to use Taylor-approximations of the dynamics.

The inherent disadvantage of the approach described in \autoref{text:roa_est} lies in the probabilistic character of the estimate. 
Especially for high dimensional systems, one would have to simulate the dynamics for a plethora of initial states in order for \autoref{alg:tvlqr} to yield dependable results.
Depending on the system, performing such exhaustive simulations of multi-body dynamics could be a time-consuming undertaking.
The possibilities for parallelization exist, but are limited due to the sequential nature of \autoref{alg:tvlqr}. 
One could carry out the sampling and simulation in parallel, join processes and only consider the failed simulation with the smallest initial cost-to-go, when shrinking funnels.
While this approach would certainly speed up computations, it will most likely not be able to fully alleviate the \textit{curse of dimensionality}. 
In order to increase the likelihood of finding initial states that after simulation of $\mathbf{f}_{cl}$ lead outside $\mathcal{B}_f$, one could add heuristics to the sampling process.
For example, one could modify the \textit{direct-sphere} algorithm, such that samples are preferably drawn from the outer regions of $\mathcal{B}_0$.
In doing so, the likelihood of the simulation failing and therefore enhancing the estimate are increased. 
A similar approach was found to increase the performance of probabilistic RoA estimation for linear time invariant dynamics \citep{najafi2016}.
Furthermore, in order to obtain funnels that reach towards the end of $\mathbf{x}^*, \mathbf{u}^*$, one could perform the piecewise simulations in reverse direction, i.e. first sample from the penultimate ellipsoid and simulate until the last, then move from back to front.
A combination of forward and backward piecewise simulations could be viable. One could perform a couple of "forward" runs to quickly shrink the funnel and then continue with the reverse piecewise simulation in order to make sure, it extends until $\mathcal{B}_f$. 
However, such an approach would make it harder to include certain constraints, e.g. integral ones, such as the fuel constraint in \textit{scenario I}.

In \textit{scenario I}, while the RoA estimations converged and were plausible with respect to the fuel constraint (\autoref{eq:fuel_constr}), there is some remaining uncertainty in the final estimates.
Different seeding of the random number generator leads to results that are close to one-another, but not equal (\autoref{fig:funnels_alpha} a).
An explanation as to why different estimation processes do not yield more similar results is, that as $\mathcal{B}_0$ shrinks, it contains mostly initial states that are part of the actual RoA.
Hence, it becomes increasingly more unlikely for a simulation to fail, and no information is added to the estimate.
Even worse, in case of high dimensional state spaces, the \textit{curse of dimensionality} affects the estimation process twice: not only does randomly picking an initial state that after simulation leads outside $\mathcal{B}_f$ take longer than for low dimensional systems, but also the simulation itself is more demanding.
Furthermore, the fuel constraint seems to negatively affect the convergence of \autoref{alg:tvlqr}.
This can be seen in \autoref{fig:funnels_alpha} a), in which the final estimates much more agree with one another, when the fuel constraint is bypassed, i.e. $\alpha=\infty$. 

In \textit{scenario II}, a precomputed nominal trajectory $\mathbf{x}^*,\mathbf{u}^*$ could successfully be stabilized using TVLQR.
The estimated RoA could partially be validated in lab experiments.
However, certain initial states estimated to be part of $\mathcal{B}_0$ lead outside of $\mathcal{B}_f$ due to exceeding the terminal velocity limit.
Most likely, this was caused by the actuation deadband of the thrusters, that was not considered in trajectory optimization and can not properly be addressed by a linear controller, such as TVLQR.
Recall that towards the end of $\mathbf{x}^*$, the freeflyer was supposed to break and hold the position.
When close to the holding pose, such that $|v_x|=|v_y| \approx 0$, TVLQR demands forces that fall below the lower threshold of minimal actuation of the thrusters. 
This results in early abortion of breaking and subsequent overshooting.
Another effect of the deadband is a limit cycle that occurs during station holding.
For trajectory optimization in \textit{scenario II}, one could opt for a higher dimensional $\mathbf{u} \in \mathbb{R}^8$ that consists of actuator forces rather than generalized forces.
Doing so would allow to include the actuation deadband within trajectory optimization, which could lead to trajectories in which opposing thrusters are active during station holding in order to prevent a nonzero terminal velocity.
Besides, in order to enable freeflyers to operate in low-thrust regimes and to reduce steady state error, sigma-delta modulation could be used \citep{zappulla2017sigmadelta}.

%% file: sections/conclusion.tex
\label{text:conclusion}
In this work, a probabilistic method for computing the funnel around trajectories based on~\citep{reist2010} has been conceived and applied to two free floating systems under TVLQR control.
\autoref{text:background} introduces the free floating dynamics and the notion of funnels based on Lyapunov theory.
In \autoref{text:stabilization} describes, how a precomputed nominal trajectory can be stabilized with a time varying linear control policy (TVLQR), that facilitates usage of Lyapunov based methods for stability analysis.
In \autoref{text:roa_est}, a sampling based method for estimating a probabilistic funnel around a nominal trajectory is conceived and applied to two distinct scenarios in \autoref{text:results}.
In \textit{scenario I}, we consider the detumble phase of a simulated ADR scenario consisting of two satellites connected by a 3R robotic manipulator, whereas for \textit{scenario II}, lab experiments onboard ELISSA have been conducted considering in-plane motion of a sigle freeflyer. 
Reflecting on \autoref{text:results} and \autoref{text:discussion}, one can say, that \textit{scenario I} demonstrates the compromises of probabilistic RoA estimation and problems related to dimensionality that are yet to be solved, while \textit{scenario II} highlights the importance of proper and accurate modelling, when planning to rely on RoA estimates in real life (space) applications. 

As stated in \autoref{text:intro}, for the control of future autonomous spacecraft during proximity operations, the concept of funnels is very interesting for ensuring the feasibility and composability of (sub-) maneuvers.
In particular, probabilistic estimation methods allow for modelling a wide range of constraints and effects, that need to be taken into consideration, when the goal is to reason about real-world scenarios.